\documentclass[journal]{IEEEtran}

\pdfoutput=1

\usepackage{amsmath}
\usepackage{amssymb}
\usepackage{amsfonts}
\usepackage{amscd}
\usepackage{amsthm}
\usepackage[noadjust]{cite}
\usepackage{threeparttable}
\usepackage{textcomp}
\usepackage{multirow}

\usepackage{mathrsfs}
\usepackage{bbm}
\usepackage{amsbsy}

\usepackage{psfrag}

\usepackage{paralist}

\usepackage{xfrac}

\newtheorem{theorem}{Theorem}
\newtheorem{lemma}{Lemma}

\theoremstyle{definition}
\newtheorem{example}{Example}
\newtheorem{definition}{Definition}

\makeatletter
\renewcommand*\env@matrix[1][c]{\hskip -\arraycolsep
  \let\@ifnextchar\new@ifnextchar
  \array{*\c@MaxMatrixCols #1}}
\makeatother
\usepackage{graphicx}

\newcommand{\Fb}{\mathbbmss{F}}
\newcommand{\Zb}{\mathbbmss{Z}}

\newcommand{\zb}{\pmb{z}}
\newcommand{\yb}{\pmb{y}}
\newcommand{\cb}{\pmb{c}}
\newcommand{\wb}{\pmb{w}}
\newcommand{\Cs}{\mathscr{C}}
\newcommand{\Gb}{\pmb{G}}
\newcommand{\gb}{\pmb{g}}
\newcommand{\mmod}{{\rm~mod~}}
\newcommand{\dmin}{d}
\newcommand{\Csout}{\mathscr{C}^{\text{(out)}}}
\newcommand{\nout}{n^{\text{(out)}}}

\newcommand{\nin}{n^{\text{(in)}}}
\newcommand{\din}{d^{\text{(in)}}}
\newcommand{\Csin}{\mathscr{C}^{\text{(in)}}}

\title{New Error Correcting Codes for Informed Receivers}
\author{
\IEEEauthorblockN{Lakshmi~Natarajan, Yi~Hong, and Emanuele~Viterbo}%
}

\begin{document}

\maketitle

\begin{abstract}
We construct error correcting codes for jointly transmitting a finite set of independent messages to an \emph{informed receiver} which has prior knowledge of the values of some subset of the messages as side information. The transmitter is oblivious to the message subset already known to the receiver and performs encoding in such a way that any possible side information can be used efficiently at the decoder.
We construct and identify several families of algebraic error correcting codes for this problem using cyclic and maximum distance separable (MDS) codes. The proposed codes are of short block length, many of them provide optimum or near-optimum error correction capabilities and guarantee larger minimum distances than known codes of similar parameters for informed receivers.
The constructed codes are also useful as error correcting codes for index coding when the transmitter does not know the side information available at the receivers.
\end{abstract}


\section{Introduction} \label{sec:1}

We consider the channel coding problem where the transmitter jointly encodes a set of $L$ independent messages while the receiver has prior knowledge of the values of some subset of the $L$ messages. The transmitter is ignorant of the subset of source messages already known at the receiver, and hence, is required to encode the messages in such a way that every possible side information at the receiver can be exploited efficiently.
Following~\cite{DBT_ISIT_01}, we refer to this communication problem as \emph{coding for informed receiver}.
An equivalent communication scenario is the broadcast of $L$ messages to multiple receivers where each receiver has side information of a different subset of source messages.
Applications of this problem include the broadcast phase of decode-and-forward protocol in multi-way relay networks~\cite{XFKC_CISS_06,XuS_ITW_07}, and retransmissions in a broadcast channel where each receiver has successfully decoded some subset of the messages from previous transmissions~\cite{BiK_INFOCOM_98,Met_Comm_84}. 
It is known that linear coding schemes for informed receivers involve the design of $L$ codes $\Cs_1,\dots,\Cs_L$ which are linearly independent (as vector spaces) such that the sum of any subset of the $L$ codes is a good error correcting code~\cite{XFKC_CISS_06,DBT_ISIT_01}.
To the best of our knowledge, only a few explicit constructions of codes for informed receivers based on convolutional codes~\cite{XFKC_IT_07,MLV_PIMRC_12}, LDPC codes~\cite{KeK_ISIT_10}, and errors-and-erasures decoding of linear codes~\cite{DBT_ISIT_01} are available.

\let\thefootnote\relax\footnotetext{
The authors are with the Department of Electrical and Computer Systems Engineering, Monash University, VIC 3800, Australia (e-mail: \{lakshmi.natarajan, yi.hong, emanuele.viterbo\}@monash.edu). This work was supported by the Australian Research Council under Grant Discovery Project No.~DP160101077.
 
\copyright 2016 IEEE. Personal use of this material is permitted. Permission from IEEE must be obtained for all other uses, in any current or future media, including reprinting/republishing this material for advertising or promotional purposes, creating new collective works, for resale or redistribution to servers or lists, or reuse of any copyrighted component of this work in other works.
} 

In this paper, we construct several families of linear block codes for informed receivers by using algebraic error correcting codes. The constructed coding schemes are of modest block lengths, most of the presented schemes provide optimum or near-optimum error correction capability and guarantee larger minimum distances than known coding schemes of similar rate and block length available from~\cite{DBT_ISIT_01}.
After illustrating the problem with an example using a new optimal binary code for informed receivers (Section~\ref{sec:ECCIR}), we characterize the family of maximal distance separable (MDS) codes for this channel (Section~\ref{sec:MDS_for_informed_receivers}). We then construct and identify several families of binary codes for informed receivers using cyclic codes, quadratic and cubic residue codes, and concatenated coding (Section~\ref{sec:binary_codes}).

A related problem is that of \emph{index coding}~\cite{YBJK_IEEE_IT_11} where $L$ messages are to be broadcast to a set of receivers and each receiver demands some subset of the source messages while having prior knowledge of a different subset as side information. The error correcting codes for this problem available in~\cite{DSC_IT_13,ThR_ISIT_15} assume that the demands and side information of the receivers are known to the transmitter. On the other hand, the codes of this paper are suitable when no such knowledge is available at the encoder.

The codes $\Cs$ constructed in this paper decompose as a direct sum of $L$ subcodes $\Cs_1,\dots,\Cs_L$ such that each of the $2^L-1$ subcodes of $\Cs$ formed as the sum of some subset of $\{\Cs_1,\dots,\Cs_L\}$ is a good error correcting code for its rate and blocklength. 
Constructions of pairs \mbox{$\Cs \supset \Cs'$}, or even chains $\Cs \supset \Cs' \supset \Cs'' \supset \cdots$, of linear codes have been previously investigated in the literature; see, for example,~\cite{XFKC_IT_07,LVC_IT_10,BaC_ITW_11}.
However, these nested codes are not useful when the receiver side information is an arbitrary subset of source messages.

{\it Notation:} 
Matrices and row vectors are denoted by bold upper and lower case letters, respectively.
The minimum Hamming distance of a code $\Cs$ is denoted by $\dmin(\Cs)$. The symbol $d^*(n,k)$ denotes the maximum of the minimum Hamming distances over all $[n,k]$ \emph{binary} linear codes. Unless otherwise stated all the values of, and bounds on, $d^*(n,k)$ are referenced from the table of best known linear codes available in~\cite{Gra_codetables}.

\section{Review of linear and cyclic codes} \label{sec:linear_cyclic_codes} 

Let $q$ be a prime power and $\Fb_q$ be the finite field of size $q$. 
In order to design coding schemes for receivers with side information, we will consider finite collections of length $n$ linear codes over $\Fb_q$.
A collection $\{\Cs_1,\dots,\Cs_L\}$ of linear codes is \emph{linearly independent} if the only choice of $\cb_\ell \in \Cs_\ell$, $\ell=1,\dots,L$, satisfying $\sum_{\ell=1}^{L}\cb_\ell=\pmb{0}$ is $\cb_1=\cdots=\cb_L=\pmb{0}$. If $\Cs_1,\dots,\Cs_L$ are linearly independent, the sum code $\Cs=\sum_{\ell=1}^{L}\Cs_\ell$ is their direct sum as a vector space.

We now introduce the notation and briefly review some of the relevant properties of cyclic codes based on~\cite{MaS_North_88,Ple_Wiley_89}.
We consider cyclic codes of length $n$ over $\Fb_q$ with \mbox{$\gcd(n,q)=1$}. Label the coordinates of \mbox{$\cb \in \Fb_q^n$} with the elements of \mbox{$\Zb_n=\{0,1,\dots,n-1\}$} and associate the vector $\cb=(c_0,\dots,c_{n-1})$ with the polynomial $c(x)=c_0+c_1x + \cdots + c_{n-1}x^{n-1}$. With this correspondence a \emph{cyclic code} $\Cs$ is an ideal in the ring $R_n=\Fb_q[x]/(x^n-1)$. 
We use $g(x)$ to denote the {generator polynomial} of $\Cs$ and $h(x)=(x^n-1)/g(x)$ to denote its {check polynomial}.


The \emph{$q$-cyclotomic coset modulo $n$} of \mbox{$i \in \Zb_n$} is the set \mbox{$C_i=\{i,qi,q^2i,\dots\}$}, where the arithmetic is performed modulo $n$. Let $\alpha$ be a primitive $n^{\text{th}}$ root of unity in some extension field \mbox{$\Fb_{q^m} \supset \Fb_q$}. Then, for some \mbox{$T \subset \Zb_n$}, we have 
$h(x)=\prod_{i \in T}(x - \alpha^i)$ and $g(x)=\prod_{i \in \overline{T}} (x-\alpha^i)$,
where $\overline{T}$ is the complement of $T$ in $\Zb_n$. The sets $T$ and $\overline{T}$ are the \emph{non-zeroes} and \emph{zeroes} of $\Cs$, respectively, and each is a union of $q$-cyclotomic cosets modulo $n$. The dimension of $\Cs$ is $|T|$.
When \mbox{$q=2$}, the subcode of $\Cs$ consisting of all even weight codewords of $\Cs$ is a cyclic code with non-zeroes $T\setminus\{0\}$.
A cyclic code is \emph{irreducible} if it contains no non-trivial cyclic subcodes. A cyclic code with non-zeroes $T$ is irreducible if and only if $T$ is a cyclotomic coset by itself.

If $T_1,\dots,T_L$ are the non-zeroes of cyclic codes $\Cs_1,\dots,\Cs_L$ then $\cup_{\ell=1}^{L}T_\ell$ is the set of non-zeroes of $\Cs=\sum_{\ell=1}^{L}\Cs_\ell$.
With the notation as above, we state the following fact whose proof is straightforward.
\begin{lemma} \label{lem:nonzeroes_partition}
The collection $\{\Cs_1,\dots,\Cs_L\}$ is linearly independent if and only if $T_1,\dots,T_L$ are non-intersecting. 
\end{lemma}

Let $a$ be any integer with \mbox{$\gcd(a,n)=1$}. The function 
\mbox{$i \to ai \mmod n$},
is a permutation on the set of coordinates $\Zb_n$ since $a$ has a multiplicative inverse in $\Zb_n$. The polynomial $\mu_a\left(c(x)\right)$ obtained by applying this permutation on the coordinates of $c(x)$ is 
\begin{align*}
c(x^a) = c_0 + c_1x^a + \cdots c_{n-1}x^{a(n-1)} \mmod (x^n-1).
\end{align*} 
When applied to a cyclic code (ideal) $\Cs \subset R_n$, $\mu_a(\Cs)$ is the set of all polynomials $\mu_a(c(x))$ with $c(x) \in \Cs$, and further, $\mu_a(\Cs)$ is itself a cyclic code.
\begin{lemma} \label{lem:nonzeroes_aT}
The set of non-zeroes of $\Cs$ is $T$ if and only if the set of non-zeroes of $\mu_a(\Cs)$ is $a^{-1}T$.
\end{lemma}
\begin{IEEEproof}
An element $j \in \Zb_n$ is a non-zero of a cyclic code if and only if there exists a codeword polynomial for which $\alpha^j$ is not a root. 
If \mbox{$j \in T$}, then there exists $c(x) \in \Cs$ such that $c(\alpha^j) \neq 0$. 
The evaluation of $c'(x)=c(x^a) \in \mu_a(\Cs)$ at $x=\alpha^{a^{-1}j}$ yields $c'(\alpha^{a^{-1}j})=c(\alpha^j) \neq 0$ showing that $a^{-1}j$ is a non-zero of $\mu_a(\Cs)$. The proof of converse is similar.
\end{IEEEproof}


\section{Error Correction for Informed Receivers} \label{sec:ECCIR}
 
Consider a vector $\wb$ of length $kL$ over $\Fb_q$ composed of $L$ independent message vectors \mbox{$\wb_1,\dots,\wb_L \in \Fb_q^k$}, i.e., \mbox{$\wb=\left(\wb_1,\dots,\wb_L\right)$}. The message $\wb$ is encoded by a linear code $\Cs$ (not necessarily cyclic) of length $n$ using a generator matrix \mbox{$\Gb \in \Fb_q^{kL \times n}$} of full-rank $kL$. Let \mbox{$\Gb_1,\dots,\Gb_L \in \Fb_q^{k \times n}$} be the submatrices of $\Gb$ corresponding to $\wb_1,\dots,\wb_L$, respectively, i.e.,
 $\Gb = \begin{pmatrix} \Gb_1^\intercal~\cdots~\Gb_L^\intercal \end{pmatrix}^\intercal$.
The message $\wb$ is encoded into the length $n$ codeword
$\cb=\wb\Gb=\sum_{\ell=1}^{L}\wb_\ell\Gb_\ell$. 
Using $\Cs_\ell$ to denote the linear code with generator matrix $\Gb_\ell$, we observe that $\Cs_1,\dots,\Cs_L$ are linearly independent and $\Cs$ is their direct sum.

For \mbox{$S \subsetneq \{1,\dots,L\}$}, consider the receiver ${\sf Rx}_S$ that has prior information of the values of $\wb_\ell$, \mbox{$\ell \in S$}. 
Note that this includes the case \mbox{$S=\varnothing$}, i.e., no side information.
On observing the channel output $\yb=\cb + \zb$, where $\zb$ is the error vector, ${\sf Rx}_S$ removes the contributions of $\wb_\ell$, \mbox{$\ell \in S$}, from $\yb$ to arrive at
\begin{equation*}
 \yb_S = \yb - \sum_{\ell \in S} \wb_\ell\Gb_\ell = \sum_{\ell \in \bar{S}} \wb_\ell\Gb_\ell + \zb,
\end{equation*} 
where $\bar{S}$ is the complement of $S$ in $\{1,\dots,L\}$. The unknown messages $\wb_\ell$, $\ell \in \bar{S}$, are then estimated by decoding $\yb_S$ to 
\begin{equation*}
\Cs_{\bar{S}} = \sum_{\ell \in \bar{S}} \Cs_\ell = \Big\{ \sum_{\ell \in \bar{S}}\wb_\ell\Gb_\ell \, \Big\vert \, \wb_\ell \in \Fb_q^{k}, \ell \in \bar{S} \, \Big\}.
\end{equation*} 
Note that $\Cs_{\bar{S}}$ is a subcode of $\Cs$ of dimension $k|\bar{S}|$.
In order to maximize the error correction capability at ${\sf Rx}_S$ we require that $\Cs_{\bar{S}}$ be a good linear error correcting code. 

We are interested in the scenario where the transmitter is oblivious to the side information $S$ available at the receiver, and thus, we require that each of the $2^L-1$ subcodes $\Cs_{\bar{S}}$ of the code $\Cs$, one corresponding to each possible side information configuration $S \subsetneq \{1,\dots,L\}$, be a good error correcting code, i.e., with a large minimum Hamming distance $\dmin(\Cs_{\bar{S}})$. 

\begin{definition}
An \emph{error correcting code for informed receivers (ECCIR)} encoding $L$ messages is a linearly independent collection $\{\Cs_1,\dots,\Cs_L\}$ of $L$ linear codes.
\end{definition}

The code design objective is to construct $\{\Cs_1,\dots,\Cs_L\}$ such that all the minimum Hamming distances $d(\Cs_{\bar{S}})$, for every \mbox{$S \subsetneq \{1,\dots,L\}$}, are as large as possible.



\begin{example} \label{ex:binary_ECCIR}
\emph{A new optimum binary ECCIR of length \mbox{$n=31$} for \mbox{$L=3$} messages of size \mbox{$k=10$} each}:
Consider cyclic codes of length \mbox{$n=31$} over the binary alphabet \mbox{$q=2$}. 
The cyclotomic cosets are
\begin{IEEEeqnarray}{l'l}
 C_1 =\{1,2,4,8,16\}, & C_3 =\{3,6,12,24,17\} \IEEEnonumber \\
 C_5 =\{5,10,20,9,18\},  & C_7=\{7,14,28,25,19\} \IEEEnonumber \\
 C_{11} = \{11,22,13,26,21\},  & C_{15}=\{15,30,29,27,23\},\IEEEnonumber
\end{IEEEeqnarray} 
and $C_0=\{0\}$. The $[31,30,2]$ single-parity check code $\Cs$ is the cyclic code with non-zeroes $T=\{1,\dots,30\}$. Consider the codes $\Cs_\ell$, $\ell=1,2,3$, with non-zeroes $T_1=C_1 \cup C_3$, $T_2=C_5 \cup C_{15}$ and $T_3=C_7 \cup C_{11}$, respectively. Since $T_1 \cup T_2 \cup T_3$ is a partition of $T$, $\Cs$ is a direct sum of $\Cs_\ell$, $\ell=1,2,3$ (from Lemma~\ref{lem:nonzeroes_partition}). 

\emph{Equivalence of codes:} Since \mbox{$n=31$} is prime, \mbox{$\gcd(a,n)=31$} for any non-zero \mbox{$a \in \Zb_{31}$}.
Observe that \mbox{$T_1=25T_2=9T_3$}, and \mbox{$25^{-1}=5$} and \mbox{$9^{-1}=7$} in $\Zb_{31}$. Using Lemma~\ref{lem:nonzeroes_aT}, we deduce that \mbox{$\Cs_1=\mu_5(\Cs_2)=\mu_7(\Cs_3)$}. It follows that $\Cs_1,\Cs_2,\Cs_3$ are equivalent up to coordinate permutations, and in particular, they have the same minimum distance. Similarly, since $T_1\cup T_2=9(T_1\cup T_3)=25(T_2 \cup T_3)$, the three codes $\Cs_1 + \Cs_2$, $\Cs_1 + \Cs_3$ and $\Cs_2 + \Cs_3$ are equivalent.

\emph{Minimum distance of $\Cs_2 + \Cs_3$:}
The code $\Cs_2 + \Cs_3$ (zeroes $C_0 \cup C_1 \cup C_3$) is the even weight subcode of the double-error correcting BCH code (zeroes $C_1 \cup C_3$) with parameters $[31,21,5]$. Hence, $\dmin(\Cs_2 + \Cs_3) \geq 6$. Since \mbox{$d^*(31,20)=6$}, we conclude that $\Cs_2+\Cs_3$ is a $[31,20,6]$ code, where $d^*(n,k)$ is the maximum of the minimum Hamming distances over all $[n,k]$ binary codes.

\emph{Minimum distance of $\Cs_1$:}
The code $\Cs_1$ is equivalent to $\mu_{-1}(\Cs_1)$ (non-zeroes $C_{-1} \cup C_{-3}$). 
Note that \mbox{$-1=30$} and \mbox{$-3=28$} in $\Zb_{31}$.
The dual of $\mu_{-1}(\Cs_1)$ has zeroes at $C_1 \cup C_3$, and hence, $\left(\mu_{-1}(\Cs_1)\right)^\perp$ is the double-error correcting primitive BCH code with parameters \mbox{$[31,21,5]$}. 
Using the Carlitz-Uchiyama bound~\cite[p.~280]{MaS_North_88}, we know that the minimum distance of the dual of the $[31,21,5]$ primitive BCH code is even and satisfies the lower bound
 $d\left( \mu_{-1}(\Cs_1) \right) \geq 2^{(5-1)} - 2^{\sfrac{5}{2}}=10.34..$,
i.e., $\dmin(\mu_{-1}(\Cs_1)) \geq 12$. Using the fact that $d^*(31,10)=12$, we conclude that $\mu_{-1}(\Cs_1)$ and $\Cs_1$ are $[31,10,12]$ codes. 

In summary, $\Cs_1$, $\Cs_2$ and $\Cs_3$ are all $[31,10,12]$ codes, the sum of any two of these three codes is $[31,20,6]$, and the sum of all three is the $[31,30,2]$ code. 
All these codes have the optimum minimum distance for their length and dimension. 
At the receiver, the minimum distances $\dmin(\Cs_{\bar{S}})$ corresponding to the side information configuration $S$, with  $|S|=0,1,2$, are $2,6,12$, respectively.
\hfill\IEEEQED
\end{example}

\section{Maximum Distance Separable Codes} \label{sec:MDS_for_informed_receivers}

In this section we construct ECCIRs such that all the \mbox{$2^L-1$} codes $\Cs_{\bar{S}}$, $S \subsetneq \{1,\dots,L\}$, meet their respective Singleton bounds 
\mbox{$\dmin(\Cs_{\bar{S}}) \leq n - k|\bar{S}| + 1$}.

\begin{definition}
A collection of codes $\{\Cs_1,\dots,\Cs_L\}$ is \emph{maximum distance separable for informed receivers (MDSIR)} if \mbox{$\dmin(\Cs_{\bar{S}})=n-k|\bar{S}|+1$} for every $S \subsetneq \{1,\dots,L\}$.
\end{definition}

We construct MDSIR codes $\{\Cs_1,\dots,\Cs_L\}$ for \mbox{$k=1$} and arbitrary $L$, i.e., where each message $\wb_\ell$ is a scalar by itself. 
For any $k_0$ and $L_0$ with \mbox{$k_0L_0=L$}, an MDSIR code $\{\Cs_1',\dots,\Cs_{L_0}'\}$ for $L_0$ messages of length $k_0$ each can be readily obtained from $\{\Cs_1,\dots,\Cs_L\}$ by setting \mbox{$\Cs_m'=\sum_{\ell=(m-1)k_0+1}^{mk_0}\Cs_\ell$}.

The generator matrix $\Gb_\ell$ of $\Cs_\ell$, $\ell=1,\dots,L$, consists of a single vector $\gb_\ell \in \Fb_q^n$, and the generator $\Gb \in \Fb_q^{L \times n}$ of $\Cs=\sum_{\ell=1}^{L}\Cs_\ell$ consists of rows $\gb_1,\dots,\gb_L$.
The generator $\Gb_{\bar{S}}$ of $\Cs_{\bar{S}}$ is a submatrix of $\Gb$ composed of the rows $\gb_\ell$, $\ell \in \bar{S}$. 
The following lemma characterizes the generator matrices $\Gb$ of MDSIR codes for informed receivers when $k=1$.

\begin{lemma} \label{lem:MDSIR_code}
The matrix $\Gb \in \Fb_q^{L \times n}$ is the generator of an MDSIR code if and only if every square submatrix of $\Gb$ is nonsingular.
\end{lemma}
\begin{IEEEproof}
We know that $\Cs_{\bar{S}}$ is MDS, i.e., 
$\dmin(\Cs_{\bar{S}})=n-|\bar{S}|+1$, 
if and only if every \mbox{$|\bar{S}| \times |\bar{S}|$} submatrix of $\Gb_{\bar{S}}$ is nonsingular.
Equivalently, $\Cs_{\bar{S}}$ is MDS if and only if any square submatrix of $\Gb$ obtained by selecting the rows corresponding to $\bar{S}$ and any $|\bar{S}|$ columns is nonsingular. By letting $S$ vary over all subsets of $\{1,\dots,L\}$ we arrive at the statement of the lemma.
\end{IEEEproof}

Matrices with every square submatrix being nonsingular are known to be related to the generator matrices of (traditional) MDS codes~\cite{MaS_North_88}.
Let $\pmb{A}=[\pmb{I} \, \vert \, \Gb]$ be the systematic generator matrix of an $[n+L,L]$ linear code over $\Fb_q$. Then we have 

\begin{theorem}[{\cite[p.~321]{MaS_North_88}}] \label{thm:nonsingular_submatrix}
Every square submatrix of $\Gb$ is nonsingular if and only if $\pmb{A}=[\pmb{I} \, \vert \, \Gb]$ generates an MDS code.
\end{theorem}

It follows that we can construct length $n$ MDSIR codes for $L$ symbols by puncturing the information coordinates of any $[n+L,L]$ systematic MDS code, such as the systematic versions of extended Reed-Solomon (RS) and generalized RS codes. 
For example, a length $n$ MDSIR code for $L$ messages and \mbox{$k=1$} exists over all $\Fb_q$ with \mbox{$q > n+L$} since an \mbox{$[n+L,L,n+1]$} generalized RS code exists over such $\Fb_q$.

\subsection*{Comparison with the ECCIRs of~\cite{DBT_ISIT_01}}

A construction of ECCIRs similar to that of this section was proposed in~\cite{DBT_ISIT_01} using an approach based on errors-and-erasures decoding of linear codes.
For any $q,k$, \cite{DBT_ISIT_01} shows that if \mbox{$\pmb{A}=[\pmb{I}\,\vert\,\Gb]$} is the generator of an \mbox{$[n+kL,kL,d]$} code (not necessarily MDS) over $\Fb_q$, then the $L$ submatrices $\Gb_1,\dots,\Gb_L$ of $\Gb=(\Gb_1^\intercal,\dots,\Gb_L^\intercal)^\intercal$ generate an ECCIR with \mbox{$\dmin(\Cs_{\bar{S}}) \geq \max\{d - k|\bar{S}|,0\}$}. 
We remark that if $\pmb{A}$ generates an MDS code, i.e., if $d=n+1$, the construction of~\cite{DBT_ISIT_01} yields an MDSIR code.
Compared to~\cite{DBT_ISIT_01}, our approach illuminates the direct and strong relation between MDS and MDSIR codes (Lemma~\ref{lem:MDSIR_code} and Theorem~\ref{thm:nonsingular_submatrix}).

As illustrated by the following example, the binary non-MDSIR codes constructed in this paper (Example~\ref{ex:binary_ECCIR} and Section~\ref{sec:binary_codes}) can guarantee larger minimum distances than the binary codes constructed using the technique of~\cite{DBT_ISIT_01}.

\begin{example}
To generate an ECCIR with the same parameters (\mbox{$k=10$}, \mbox{$L=3$}, \mbox{$n=31$}, \mbox{$q=2$}) as the new code of Example~\ref{ex:binary_ECCIR}, the approach of~\cite{DBT_ISIT_01} starts with the best known binary code of length \mbox{$n+kL=61$} and dimension \mbox{$kL=30$}, which has minimum distance \mbox{$d=12$}~\cite{Gra_codetables}. For $|S|=0,1,2$, this technique guarantees $\dmin(\Cs_{\bar{S}}) \geq 0,0,2$, respectively. 
While the bounds for $|S|=0,1$ are trivial, the bound for $|S|=2$ is significantly lower than $\dmin(\Cs_{\bar{S}})=12$ achieved in Example~\ref{ex:binary_ECCIR}.
\hfill\IEEEQED
\end{example}

\section{Binary Codes for Informed Receivers} \label{sec:binary_codes}

In this section we construct and identify several families of binary ECCIRs using cyclic codes and the MDSIR codes of Section~\ref{sec:MDS_for_informed_receivers}.

\subsection{Code Concatenation}

Binary ECCIRs can be obtained from the MDSIR codes of Section~\ref{sec:MDS_for_informed_receivers} by concatenating them with a binary inner code.
Let \mbox{$q=2^k$} and let $\{\Csout_1,\dots,\Csout_L\}$ be a length $\nout$ MDSIR code for $L$ symbols over $\Fb_q$, i.e., each $\Csout_\ell$ is of dimension $1$ over $\Fb_q$, or equivalently, dimension $k$ over $\Fb_2$. Each of the $\nout$ $\Fb_{2^k}$-symbols of the outer MDSIR code $\Csout=\sum_{\ell=1}^{L}\Csout_\ell$ is linearly mapped to a length $k$ binary vector and then encoded by an $[\nin,k,\din]$ binary inner code. The resulting binary ECCIR $\Cs_1,\dots,\Cs_L$ encodes $L$ binary messages of size $k$ each into a length $n=\nout\nin$ codeword over $\Fb_2$. Suppose a receiver ${\sf Rx}_S$ has prior knowledge of the messages with indices in $S \subsetneq \{1,\dots,L\}$. The effective binary code $\Cs_{\bar{S}} \subset \Fb_2^n$ at this receiver is the concatenation of $\Csout_{\bar{S}}=\sum_{\ell \in \bar{S}}\Csout_{\ell}$ and the $[\nin,k,\din]$ binary inner code. Since $\dmin(\Csout_{\bar{S}})=\nout-|\bar{S}|+1$, we have
\begin{equation} \label{eq:dmin_concat}
 \dmin(\Cs_{\bar{S}}) \geq \din \left( \nout - |\bar{S}| + 1 \right).
\end{equation} 
The outer MDSIR code ensures that the lower bound on distance improves with the amount of side information available at the receiver.

\begin{example}
Consider $L=2$ binary messages of size $k=3$ each. Let the outer MDSIR code be of length $\nout=3$ over $\Fb_{2^k}=\Fb_8$. 
Since \mbox{$2^k>\nout+L$} such a code exists and can be constructed from a generalized RS code (Section~\ref{sec:MDS_for_informed_receivers}). 
Let the binary inner code be $[\nin,k,\din]=[7,3,4]$. The resulting binary ECCIR $\{\Cs_1,\Cs_2\}$ has length \mbox{$n=21$}. From~\eqref{eq:dmin_concat}, $\Cs_1$, $\Cs_2$ and $\Cs=\Cs_1+\Cs_2$ have dimensions $3$, $3$ and $6$, and minimum distances at least $12$, $12$ and $8$, respectively. Since $d^*(21,3)=12$ and $d^*(21,6)=8$, we conclude that $\dmin(\Cs_1)=\dmin(\Cs_2)=12$ and $\dmin(\Cs_1+\Cs_2)=8$, i.e., all three codes possess the optimum minimum Hamming distance. \hfill\IEEEQED
\end{example}

\subsection{Code Concatenation using Piret's method} \label{sec:sub:piret}

In this subsection we consider the specific case of code concatenation where \mbox{$L=\nout=2$} and the \mbox{$[\nin,k,\din]$} binary inner code $\Csin$ is an irreducible cyclic code. 
It is known that any $[\nin,k]$ irreducible cyclic code (as an ideal in $\Fb_2[x]/(x^n-1)$) is isomorphic to the finite field \mbox{$\Fb_{2^k}$~\cite[p.~225]{MaS_North_88}}.

When \mbox{$L=\nout=2$}, the binary ECCIR \mbox{$\{\Cs_1,\Cs_2\}$} has length \mbox{$n=2\nin$}. The lower bound~\eqref{eq:dmin_concat} guarantees that $\dmin(\Cs_1),\dmin(\Cs_2) \geq 2\din$. By exploiting a known technique due to Piret~\cite{Pir_ElecLett_74}, we can optimize the outer code and guarantee that $\dmin(\Cs_1),\dmin(\Cs_2)$ is larger than $2\din$.
To do so, we restrict the $2 \times 2$ generator matrix of the outer code $\Csout=\Csout_1+\Csout_2$ to the form
\begin{equation*}
 \Gb = \begin{pmatrix} 1 & \beta \\ \beta & 1 \end{pmatrix},
\end{equation*} 
where \mbox{$\beta \neq 1$}, and we use a known finite field isomorphism \mbox{$\varphi: \Fb_{2^k} \to \Csin$}~\cite{Pir_ElecLett_74,MaS_North_88} to concatenate the outer MDSIR code with the inner irreducible cyclic code. Thus, the two component codes $\Cs_1$ and $\Cs_2$ of the binary ECCIR are
\begin{align*}
\left\{ \left(\,\varphi(a),\varphi(\beta a)\,\right) \, \vert \, a \in \Fb_{2^k} \right\} \text{ and } 
\left\{ \left(\,\varphi(\beta a),\varphi(a)\,\right) \, \vert \, a \in \Fb_{2^k} \right\} \! ,
\end{align*} 
respectively.
Note that $\Cs_1$ and $\Cs_2$ are equivalent up to coordinate permutation. 
Piret~\cite{Pir_ElecLett_74} considers codes of the same structure as $\Cs_1$ and uses a search to find the value of $\beta$ that maximizes $\dmin(\Cs_1)$; see also~\cite[p.~588]{MaS_North_88}. 
The optimal values of $\beta \neq 1$ and the resulting $\dmin(\Cs_1)$ corresponding to several binary non-primitive (\mbox{$\nin \neq 2^m-1$}) irreducible cyclic codes $\Csin$ are available in~\cite{Pir_ElecLett_74,Che_OptCod_05}.
The parameters of the resulting codes $\Cs_1$, $\Cs_2$ and $\Cs_1+\Cs_2$ are shown in Table~\ref{tb:piret_technique}.
The minimum distances $d^*(2\nin,k)$ and $d^*(2\nin,2k)$ of the optimal binary linear codes with the same length and dimension as $\Cs_\ell$, $\ell=1,2$, and $\Cs_1+\Cs_2$ are shown in parentheses. If the exact value of $d^*$ is not known best available bounds are given.
Note that since $\Gb$ is a $2 \times 2$ invertible matrix, we have
$\Cs_1 + \Cs_2 = \left\{ (\pmb{a},\pmb{b}) \, \vert \, \pmb{a},\pmb{b} \in \Csin  \right\}$.
Consequently, $\Cs_1+\Cs_2$ is a $[2\nin,2k,\din]$ code.
For every ECCIR presented in Table~\ref{tb:piret_technique} we observe that $\Cs_1$ and $\Cs_2$ equal the best known code in terms of minimum distance, and $\Cs_1+\Cs_2$ has at least half the minimum distance of the best known code.

\begin{table}
\centering
\caption{Binary ECCIRs of Section~\ref{sec:sub:piret}}
\renewcommand{\arraystretch}{1.3}
\begin{tabular} {||c|c|c||}
\hline
Irreducible & Component Codes & Sum Code \\
Cyclic Code &  of ECCIR       &          \\
$\Csin$     & $\Cs_1$ and $\Cs_2$ & $\Cs=\Cs_1 + \Cs_2$ \\
\hline
[9,6,2]   & [18,6,6]~(6)      & [18,12,2]~(4) \\
\hline
[17,8,6]   & [34,8,14]~(14)      & [34,16,6]~(8-9) \\
\hline
[21,6,8]   & [42,6,20]~(20)      & [42,12,8]~(15-16) \\
\hline
[39,12,12] & [78,12,32]~(32-33) & [78,24,12]~(22-26) \\
\hline
[41,20,10] & [82,20,26]~(26-30) & [82,40,10]~(16-20) \\
\hline
[55,20,16] & [110,20,40]~(40-44) & [110,40,16]~(24-32) \\
\hline
[65,12,26] & [130,12,56]~(56-60) & [130,24,26]~(45-51) \\
\hline
\end{tabular}
\label{tb:piret_technique}
\vspace{-3mm}
\end{table}


\subsection{Codes for $L=2$ using primitive irreducible cyclic codes} \label{sec:sub:irred_BCH}

We now design ECCIRs $\{\Cs_1,\Cs_2\}$ using irreducible binary cyclic codes $\Cs_1$ and $\Cs_2$ of \emph{primitive} length $n=2^m-1$, $m \geq 3$.
Consider the cyclotomic cosets of $1$ and $3$ in $\Zb_n$, $C_1=\{1,2,\dots,2^{m-1}\}$ and $C_3=\{3,6,\dots,3\cdot2^{m-1} \mmod n\}$. Let $\Cs_1$ and $\Cs_2$ be cyclic codes with non-zeroes $C_1$ and $C_3$, respectively. 
We observe that $\Cs_1$ and $\Cs_2$ are linearly independent (since $C_1$ and $C_3$ are non-intersecting) and they encode $k=m$ message bits each (since $|C_1|=|C_3|=m$).

We know that $\Cs_1$ is a \mbox{$[2^m-1,m,2^{m-1}]$} simplex code. If \mbox{$\gcd(3,2^m-1)=1$}, we have \mbox{$\Cs_1=\mu_3(\Cs_2)$} (from Lemma~\ref{lem:nonzeroes_aT}), and hence, $\Cs_2$ is a \mbox{$[2^m-1,m,2^{m-1}]$} code as well.
If $\gcd(3,2^m-1) \neq 1$, the irreducibility property of $\Cs_2$ can still be used to compute $\dmin(\Cs_2)$ using efficient algorithms, for details see~\cite[Ch.~2]{MaS_North_88},~\cite{Goe_IT_66} and references therein. 
For instance, the value of $\dmin(\Cs_2)$ for $m=4$, $6$ and $8$, i.e., for $n=15$, $63$ and $255$ can be computed to be $6$, $24$ and $120$, respectively. 

To analyze the minimum distance of $\Cs=\Cs_1+\Cs_2$, we observe that the equivalent code $\mu_{-1}(\Cs)$ (with non-zeroes $C_{-1} \cup C_{-3}$) is the dual of the double-error correcting primitive BCH code (with zeroes $C_1 \cup C_3$).
Applying the Carlitz-Uchiyama lower bound, we know that $\dmin(\Cs)$ is even and 
\begin{equation*}
 \dmin(\Cs) = \dmin\left( \mu_{-1}(\Cs) \right) \geq 2^{m-1} - 2^{\sfrac{m}{2}}.
\end{equation*} 

The ECCIRs $\{\Cs_1,\Cs_2\}$ of this subsection for $m \leq 8$ are summarized in Table~\ref{tb:irred_BCH}. The table also shows the values of (or the best known bounds on) $d^*(n,k)$ and $d^*(n,2k)$, which correspond to the length and dimension of the components codes $\Cs_1,\Cs_2$ and the sum code $\Cs=\Cs_1+\Cs_2$, respectively. We observe that most of the codes in Table~\ref{tb:irred_BCH} equal the best known codes in terms of minimum distance. 

\begin{table}[t!]
\centering
\caption{Binary ECCIRs for $L=2$ messages with $k=m$ and $n=2^m-1$}
\renewcommand{\arraystretch}{1.1}
{{
\begin{tabular} {||c|c|c|c|c|c|c||}
\hline
\multirow{2}{*}{$n$} & \multirow{2}{*}{$k$} & \multirow{2}{*}{$\dmin(\Cs_1)$} & \multirow{2}{*}{$\dmin(\Cs_2)$} & \multirow{2}{*}{$d^*(n,k)$} & Lower bound     & \multirow{2}{*}{$d^*(n,2k)$}  \\
    &     &                &                &            & on $\dmin(\Cs)$ &              \\
\hline
7 & 3 & 4 & 4 & 4 & 2 & 2 \\
\hline
15 & 4 & 8 & 6 & 8 & 4 & 4 \\
\hline
31 & 5 & 16 & 16 & 16 & 12 & 12 \\
\hline
63 & 6 & 32 & 24 & 32 & 24 & 24-26\\
\hline
127 & 7 & 64 & 64 & 64 & 54 & 56 \\
\hline
255 & 8 & 128 & 120 & 128 & 112 & 112-120 \\
\hline
\end{tabular}
}}
\label{tb:irred_BCH}
\end{table}

\subsection{Codes for $L=2$ from quadratic residue codes} \label{sec:sub:quad_residue}

The ECCIRs of Section~\ref{sec:sub:irred_BCH} are of low rate since the component codes $\Cs_1$ and $\Cs_2$ are irreducible cyclic codes. In this subsection we identify a class of high rate ECCIRs for $L=2$ messages where each component code is a quadratic residue (QR) code and the sum code is the $[n,n-1,2]$ single-parity check code.
Binary QR codes are a family of cyclic codes defined over prime $n$ such that \mbox{$n=\pm1 \mmod 8$}. For such $n$, $\Zb_n$ is a field and $2$ is a quadratic residue$\mmod n$, i.e., $2$ is a square in $\Zb_n$. The quadratic residues $T_1=\{a^2\,\vert\,a \in \Zb_n^*\}$ form a multiplicative subgroup of index $2$ in $\Zb_n^*$ and the non-residues $T_2=\Zb_n^*\setminus T_1$ form its coset. The QR codes $\Cs_1$ and $\Cs_2$, with non-zeroes $T_1$ and $T_2$, respectively, are equivalent, are of dimension $\sfrac{(n-1)}{2}$ and have even minimum distance of value at least $\sqrt{n}$~\cite{Ple_Wiley_89,MaS_North_88}. Since $T_1$ and $T_2$ form a partition of $\Zb_n^*$, $\{\Cs_1,\Cs_2\}$ is a binary ECCIR with $\Cs_1+\Cs_2$ being the single-parity check code. 
The QR codes $\Cs_1,\Cs_2$ for the first few values of $n$ are $[7,3,4]$, $[17,8,6]$, $[23,11,8]$, $[31,15,8]$, $[41,20,10]$, $[47,23,12]$. For each of the corresponding ECCIRs, $\Cs_1$, $\Cs_2$ and $\Cs_1+\Cs_2$ have the optimum minimum distances.

\subsection{Codes for $L=2,3$ from cubic residue codes}

The scheme of Section~\ref{sec:sub:quad_residue} can be extended to \mbox{$L=3$} using \emph{cubic residue (CR) codes}~\cite{Job_IT_92,PlR_LinAlg_88}.
Binary CR codes are defined for all prime lengths $n$ for which $3|(n-1)$ and $2$ is a cubic residue$\mmod n$. 
The set of cubic residues$\mmod n$ form a subgroup of index $3$ in $\Zb_n^*$.
Let $T_1=\{a^3\,\vert\,a \in \Zb_n\}$ be the group of cubic residues, and $T_2$ and $T_3$ be its cosets in $\Zb_n^*$. 
Let $\Cs_\ell$ be the cyclic code with non-zeroes $T_\ell$, $\ell=1,2,3$. There exists $b \in \Zb_n^*$ such that $T_1=bT_2=b^2T_3$. It follows that $\Cs_1,\Cs_2,\Cs_3$ are equivalent, and so are $\Cs_1+\Cs_2,\Cs_2+\Cs_3,\Cs_1+\Cs_3$. 
The codes $\Cs_1$ and $\Cs_1+\Cs_2$ are the even-weight CR codes of length $n$ and are of dimensions $k=\sfrac{(n-1)}{3}$ and $2k=\sfrac{2(n-1)}{3}$, respectively. 
The exact minimum distances of binary CR codes of length up to $127$ are available in~\cite{Job_IT_92}. 
Note that $\{\Cs_1,\Cs_2,\Cs_3\}$ forms an ECCIR with $\Cs=\sum_{\ell=1}^{3}\Cs_\ell$ being the $[n,n-1,2]$ code. 
Further, $\{\Cs_1,\Cs_2\}$ is an ECCIR for \mbox{$L=2$} that provides rates intermediate between the \mbox{$L=2$} codes of Sections~\ref{sec:sub:irred_BCH} and~\ref{sec:sub:quad_residue}. 
The minimum distances of the ECCIRs based on the first few binary CR codes are shown in Table~\ref{tb:cubic_residue}. The values of (or bounds on) the distance of best known linear codes of the same length and dimension are shown in parentheses. Note that all ECCIRs of Table~\ref{tb:cubic_residue} provide large minimum distances.

\begin{table}
\centering
\caption{Binary ECCIRs with $\sum_{\ell=1}^{3}\Cs_\ell=[n,n-1,2]$ code}
\renewcommand{\arraystretch}{1.3}
\begin{tabular} {||c|c|c|c||}
\hline
\multirow{2}{*}{$n$} & \multirow{2}{*}{$k$} & $\Cs_{\ell}$ & $\Cs_{\ell_1} + \Cs_{\ell_2}$ \\
                     &                      &  $\ell=1,2,3$      & $1\leq \ell_1 < \ell_2 \leq 3$ \\
\hline
31  & 10 & [31,10,10] (12) & [31,20,6] (6) \\
\hline
43  & 14 & [43,14,14] (14) & [43,28,6] (6-7) \\ 
\hline
109 & 36 & [109,36,24] (26-34) & [109,72,10] (12-16) \\
\hline
127 & 42 & [127,42,28] (32-40) & [127,84,14] (14-18) \\
\hline
\end{tabular}
\label{tb:cubic_residue}
\end{table}



\end{document}